\documentclass[12pt]{article}

\usepackage{amscd, amsmath}
\usepackage{amsthm, amssymb}
\usepackage{array}

\newcommand{\CC}{\mathbb{C}}
\newcommand{\RR}{\mathbb{R}}

\newcommand{\dd}{{\rm d}}

\newcommand{\lam}{\lambda}
\newcommand{\lamb}{\bar{\lam}}

\newcommand{\chib}{\bar{\chi}}

\newcommand{\thetab}{\bar{\theta}}

\newcommand{\zb}{\bar{z}}

\newcommand{\sib}{\bar{\sigma}}

\begin{document}

\begin{titlepage}
\title{
\vskip -70pt
\begin{flushright}
{\normalsize \ ITFA-2008-21}\\
\end{flushright}
\vskip 45pt
{\bf The quantum equivariant cohomology of toric manifolds through mirror symmetry}
}
\vspace{3cm}

\author{{J. M. Baptista} \thanks{ e-mail address:
    j.m.baptista@uva.nl}  \\
{\normalsize {\sl Institute for Theoretical Physics} \thanks{ address: Valckenierstraat 65, 1018 XE Amsterdam, The Netherlands}
} \\
{\normalsize {\sl University of Amsterdam}} 
}

\date{June 2008}

\maketitle

\thispagestyle{empty}
\vspace{2cm}
\vskip 20pt
{\centerline{{\large \bf{Abstract}}}}
Using mirror symmetry as described by Hori and Vafa, we compute the quantum equivariant cohomology ring of toric
manifolds. This ring arises naturally in topological gauged sigma-models and is related to the Hamiltonian 
Gromov-Witten invariants of the target manifold.

\vspace{.35cm}

\end{titlepage}

\section{Introduction}

Mirror symmetry was described by Hori and Vafa in \cite{H-V} as a duality between sigma-models and Landau-Ginzburg
models. In the special case where the target of the sigma-model is Calabi-Yau, this symmetry is a duality of the 
full $N=2$ supersymmetric theories, but for certain other targets, namely toric varieties and complete intersections 
of hypersurfaces in toric varieties, the symmetry still exists as a duality of the associated topological theories.
The aim of this note is to illustrate in a simple setting how Hori and Vafa's description of mirror symmetry can be 
used to study not only sigma-models but also gauged sigma-models, in our case with toric targets. Since these particular gauged 
non-linear models can be obtained as quotients of gauged linear sigma-models (with a bigger gauge group), the 
applicability of mirror symmetry is not really a surprise, for \cite{H-V} is in fact all about dualities of gauged
linear models. Nevertheless, explicit mirror predictions for gauged sigma-models do not seem to exist in the
literature, so we thought that producing one such example may be interesting.

The mathematical objects that we study here through the use of mirror symmetry are the quantum equivariant cohomology
rings of toric manifolds. In the usual non-equivariant case, the quantum cohomology rings are deformations of the de
Rham cohomology rings and encode certain genus zero Gromov-Witten invariants of the target manifold $X$. In our 
equivariant (or gauged) case, the quantum rings will be deformations of the classical equivariant cohomology rings
and will presumably encode certain genus zero Hamiltonian Gromov-Witten invariants of $X$. The latter invariants study symplectic 
manifolds equipped with hamiltonian actions and have been rigorously defined and studied in the mathematical 
literature \cite{C-G-M-S, C-S}. Loosely speaking, they are given by integrals of natural cohomology classes over the moduli space of solutions of the vortex equations with target $X$. Their ring structure, however, does not seem to have been considered so far. The Hamiltonian Gromov-Witten 
invariants can be interpreted in the language of quantum field theory as expectation values of observables in 
gauged non-linear sigma-models, in a theory very much analogous to the usual sigma-models/Gromov-Witten one \cite{Bap2}. 
The BPS states are then the classical vortex solutions, and the quantum equivariant cohomology rings are identified with the rings of local observables of the gauged topological theory.

Another version of equivariant Gromov-Witten theory has been described by Givental in \cite{Giv2}. That is a different construction from the one we are considering here. In particular the invariants in \cite{Giv2} are defined by equivariant integration over the moduli space of holomorphic curves, and so the vortex moduli spaces do not enter the picture. Recently, however, it has been proposed in \cite{Gon-Wood} that Givental's equivariant invariants are related to the weak coupling limit of the Hamiltonian (or gauged) Gromov-Witten theory, i.e. to the limit $e^2 \rightarrow 0$ of the gauged sigma-model. How this limit works precisely doesn't seem to be completely clear yet. Recall that, on the other extreme, the limit $e^2 \rightarrow \infty$ has already been studied for quite some time \cite{Ga-Sa, Hori-al}: it leads to a relation between the Hamiltonian Gromov-Witten invariants of $X$ and the usual Gromov-Witten invariants of the K\"ahler quotient of $X$ by the gauge group. 

$\ $

Coming back to our study, the main result of this note is roughly the following. Let $X$ be a $k$-dimensional toric manifold obtained as the 
K\"ahler quotient of $\CC^n$ by a linear action of the torus $T^{n-k}$ with charges $Q^a_j$, where the indices run as
$1\leq j \leq n$ and $1\leq a \leq n-k$. We assume that X is Fano, i.e. that it has positive definite first Chern class.
Then for any of the natural $T^k$-actions on $X$, the quantum equivariant
cohomology ring of this manifold is the polynomial ring
\begin{equation}
QH_T^\ast (X) \ =\ \CC [ w^1 , \ldots , w^n ]\ / \ Q(X) \ ,
\label{1.1}
\end{equation}
where $Q(X)$ is the ideal generated by the binomials
\begin{equation}
 \Big( \prod_{ \{ j: Q^a_j \geq 0 \} }  (w^j)^{Q_j^a}  \Big) - e^{-t'^a} \Big( \prod_{\{ j: Q^a_j < 0 \} }  
(w^j)^{- Q_j^a} \Big)   \label{1.2}
\end{equation}
for all $a=1, \ldots , n-k$.
Recall that the usual (non-equivariant) quantum cohomology rings of toric manifolds have been rigorously computed 
\cite{Bat, Giv, C-S}, and are
\begin{equation}
QH^\ast (X) \ =\ \CC [ w^1 , \ldots , w^n ]\ / \ ( I + Q(X)) \ ,  \nonumber
\end{equation}
where $I$ is a certain ideal of linear relations. Thus the result presented here shows a clear analogy with the
classical case, where the cohomology rings of toric manifolds are
\begin{equation}
H^\ast (X) \ =\ \CC [ w^1 , \ldots , w^n ]\ / \ ( I + J) \qquad {\rm and} \qquad 
H_{T}^\ast (X) \ =\ \CC [ w^1 , \ldots , w^n ]\ / \ J \ ,
\label{1.3}
\end{equation}
with $J$ being the classical Stanley-Reisner ideal \cite{Bat, M-P}.

$\ $

Three questions that arise directly from the work in this note are the following.
Firstly, to check the prediction of mirror symmetry by computing the quantum equivariant ring directly on the 
sigma-model side. This would presumably involve computing a few genus zero Hamiltonian Gromov-Witten invariants
of $X$. For toric $X$ we note that quite a few methods and results are already available 
\cite{C-S, Bap1, Mel-Ple}.
The second question is to go on to see whether mirror symmetry can be used to predict the Hamiltonian Gromov-Witten themselves, 
not just the resulting quantum ring. This will involve computing the correlation functions in the dual 
Landau-Ginzburg model and then mapping them to the invariants of the gauged sigma-model. Finally, the third question 
would be to generalize our computations to gauged non-abelian models with Grassmannian targets. The best shot
here is probably to use the non-abelian mirror conjecture of \cite{H-V}.

$\ $

The outline of the note is as follows. In section 2 we describe with more care the toric manifolds and actions that will 
be considered here. In section 3 we use the results of \cite{H-V} to derive the Landau-Ginzburg model that is dual to the
gauged sigma-model with target $X$. In section 4 we look at the associated gauged topological theory to recognize
the observables and the role of the gauge fields. In section 5 we finally compute the quantum equivariant 
cohomology ring, illustrating the result with the example of projective spaces.

\section{Toric manifolds and actions}

  Let $X$ be any $k$-dimensional K\"ahler toric manifold that can be realized as the K\"ahler quotient of $\CC^n$ by a 
linear action of a torus $T^{n-k}$. We will denote by $Q^a_j$ the integral weights, or charges, of this torus action, 
where the indices run as $1\leq j \leq n$ and $1\leq a \leq n-k$. Then $X$ is the space of $T^{n-k}$-orbits in a subset 
of $\CC^n$ defined by equations of the form
\begin{equation}
\mu (z) \ :=\ \sum_j Q^a_j \: |z^j|^2 \  =\  r'^a \ , \qquad z \in \CC^n \ .  
\label{2.1}
\end{equation} 
Observe that the quotient space $X(r') := \mu^{-1} (r') / T^{n-k}$ depends on the chosen value of the parameter $r' \in {\mathbb R}^{n-k}$, and only for a certain range of $r'$ will $X(r')$ be isomorphic to the initial $X$ as a complex manifold. This range can be shown to be a convex cone in the parameter space ${\mathbb R}^{n-k}$, and is called the K\"ahler cone of $X$. For $r'$ outside the K\"ahler cone the quotient $X(r')$ can be very different from $X$, and can in particular be singular or have smaller dimension. In general the parameter space ${\mathbb R}^{n-k}$ will be subdivided into a number of non-intersecting cones of different dimensions, each cone corresponding to an isomorphism class of quotients. Varying $r'$ inside each of these cones will only affect the K\"ahler metric induced on the quotient $X(r')$, not the complex manifold itself. It can then be shown that for such a variation the K\"ahler class of the induced metric depends linearly on $r'$. More precisely, for $r'$ inside the K\"ahler cone of $X$ there exists a basis $\{\eta_a \}$ of the cohomology $H^2 (X; {\mathbb Z} )$ such that the K\"ahler form $\omega_{X(r')}$ can be written as
\[
\omega_{X(r')} \ = \ \sum_a  r'^a \eta_a \ .
\]  
This shows how, inside each cone, the $r'^a$'s are effectively a parametrization of the K\"ahler class.

Another standard 
fact in toric geometry is that there exists a set of primitive vectors $v^1 , \ldots , v^n  \in \mathbb{Z}^{k}$ that span 
$\mathbb{Z}^{k}$ and satisfy the condition 
\begin{equation}
\sum_j Q_j^a \: v^j \ =\  0  
 \label{2.2}
\end{equation}
for all values of $a$. In the algebraic picture of toric manifolds these $v$'s are generators of the regular fan associated to $X$. In fact, 
given the $v$'s corresponding to $X$, we can take the $Q^a$'s to be a basis of the $(n-k)$-dimensional lattice that 
solves (\ref{2.2}). As can be easily checked, after a redefinition of the $r'^a$'s if necessary, these new $Q$'s define the 
same quotient $X$ as the old ones. The linear transformation that takes the new $Q$'s to the old ones is invertible over 
the reals but in general not over the integers. These vectors $v^1 , \ldots , v^n \in {\mathbb Z}^k$ that generate the fan encode much of the information about $X$. For example the ideal $I$ that appears in the (non-equivariant) cohomology ring of $X$ is simply defined by the linear relations
\[
\sum_{i=1}^n  \langle v^i , t \rangle \; w^i \ = 0 \ , \qquad t \in {\mathbb R}^{k}\ .
\] 

$\ $

In this paper we will be considering abelian gauged sigma-models with $X$ as target manifold. Since the models are gauged, 
to define them we need an abelian action on $X$. This action can be constructed through any abelian action on $\CC^n$ that
commutes with the $T^{n-k}$-action used in the definition of $X$, and so descends to $X$. For definiteness we will take the 
$T^k$-action on $X$ induced by the linear action of this group on $\CC^n$ with integral weights $u_j^b$, where the indices run 
as $1\leq j \leq n$ and $1\leq b \leq k$. Taking the $u$'s such that the vectors $u^1 , \ldots , u^{k}  \in \mathbb{Z}^n$ 
complete the set of $Q^a$'s to form a basis of $\RR^n$, we get at the end a hamiltonian $T^k$-action on $X$ that operates through 
holomorphic transformations and whose generic orbits have real dimension $k$.

Gauged sigma-models are theories of maps $\phi$ from a Riemann surface to the target manifold $X$ coupled to gauge fields $A$.
These gauge fields should be regarded as connections on a principal bundle over the Riemann surface, or worldsheet. If 
the classical models admit a supersymmetric extension, then in addition to $\phi$ and $A$ there will be several other bosonic
and fermionic fields. The case that concerns us are the $N=2$ gauged, supersymmetric and abelian sigma-models with target
$\CC^n$ or the toric manifold $X$. In this case both the lagrangian and the several other fields of the theory can be 
conveniently rewritten in term of superfields using the Grassmann variables $\theta^{\pm}$ and $\thetab^{\pm}$. Following
the conventions of \cite{H-V} we will need the chiral fields
\begin{equation}
\Phi   \ = \ \phi (w^\pm) + \sqrt{2}\, \theta^\alpha \, \psi_\alpha (w^\pm) + 2\, \theta^+ \theta^- \, F(w^\pm)
\label{2.3}
\end{equation}
and the twisted chiral fields
\begin{align}
Y \ &= \ y(\tilde{w}^\pm) +  \sqrt{2}\, \theta^+ \,\bar{\chi}_+ (\tilde{w}^\pm) +  \sqrt{2}\, \thetab^- \,\chi_- (\tilde{w}^\pm) +
2\, \theta^+ \thetab^- \, G(\tilde{w}^\pm)   \label{2.4}  \\
\Sigma \ &= \ \sigma (\tilde{w}^\pm)  +  i \sqrt{2}\, \theta^+ \,\lamb_+ (\tilde{w}^\pm) - i \sqrt{2} \,\thetab^- \, 
\lambda_- (\tilde{w}^\pm) +  2 \,\theta^+ \thetab^- \, [D - i (F_A)_{01}] (\tilde{w}^\pm)   \ , \nonumber
\end{align}
where one should expand the various fields as functions of the variables $w^{\pm} = x^{\pm} - i \theta^{\pm} \thetab^\pm$ and 
$\tilde{w}^{\pm} = x^{\pm} \mp i \theta^{\pm} \thetab^\pm$, with $x^\pm = x^0 \pm x^1$ being the light-cone coordinates on the 
Minkowski worldsheet. The last superfield, the superfield $\Sigma$, is not a fundamental field; it is instead the 
superfield-strength of the vector multiplet $V$ associated to the abelian connection $A$, written in the Wess-Zumino gauge. 
In terms of these
superfields the various supersymmetric lagrangians that we will use can all be written in a more compact and convenient
form.

\section{The dual gauged lagrangian}

Recall that the stated purpose of this note is to use mirror symmetry to study some basic properties of the gauged
sigma-model with toric target $X$ and group $T^k$. Our first task is therefore to write down the lagrangian of this model and, 
more importantly, the lagrangian of the mirror theory. This task is in fact quite simple given the results of \cite{H-V}. 
One starts by considering the theory before the quotients, i.e. the gauged sigma-model with target $\CC^n$ and group
$T^k \times T^{n-k}$. Its lagrangian is
\begin{align} 
L \ = \ &\int \dd^4 \theta \ \bigg\{  \sum_{j=1}^{n} \bar{\Phi}_j  \Big( \exp{(2\, Q_j^a \,V'_a)}  +  \exp{(2\, u^b_j\, V_b)}\Big) 
\Phi_j \ - \ \frac{1}{2e^2} \sum_{b=1}^{k} \Big(\bar{\Sigma}_b \, \Sigma_b \Big)   \label{3.1}\\ 
&- \frac{1}{2(e')^2} \sum_{a=1}^{n-k} \Big( \bar{\Sigma}'_a \, \Sigma'_a \Big)  \bigg\}\  
 -\ \frac{1}{2} \int \dd^2 \tilde{\theta}\ \Big( t'^a\, \Sigma'_a  + t^b\, \Sigma_b  + {\rm c.c.} \Big)  \ ,  \nonumber
\end{align}  
where $e$ and $e'$ are the gauge coupling constants corresponding to the two tori of the gauge theory, while $t$ and $t'$
are the usual complex parameters of gauged sigma-models, i.e. the complex combination $r - i \theta$ of the Fayet-Iliopoulos parameter
and the theta angle. For those less familiar with the jargon, the terms in the first and second integrals are respectively
called the D-terms and the F-terms of the lagrangian. Now, according to Hori and Vafa, the mirror of this theory is another
gauged sigma-model with twisted chiral fields $Y^1 , \ldots , Y^n $ and lagrangian
\begin{align}
L_{\rm dual} \ = \ &\int \dd^4 \theta\  \bigg\{   {\rm kinetic\ term}\ (Y^j , \bar{Y}^j)\  -\  \frac{1}{2e^2}
\sum_{b=1}^{k} \Big(\bar{\Sigma}_b\, \Sigma_b \Big)\ -\ \frac{1}{2(e')^2} \sum_{a=1}^{n-k} \Big(\bar{\Sigma}'_a \, \Sigma'_a \Big)  
\bigg\}    \label{3.2} \nonumber \\
&+\frac{1}{2} \int \dd^2 \tilde{\theta}\ \bigg\{  \Sigma'_a\: \Big(Q_j^a Y^j - t'^a \Big)\ +\  \Sigma_b \Big(u^b_j Y^j - t^b \Big) 
\ +\ \sum_{j=1}^{n} \exp{(-Y^j)} \ +\ {\rm c.c.} \bigg\}  .  \nonumber
\end{align}
In the derivation of \cite{H-V} the first two F-terms of $L_{\rm dual}$ come from the application of T-duality to $L$, 
whereas the exponential F-term is a non-perturbative contribution from instantons. A very important property of $L_{\rm dual}$
is that at a D-term level the matter and gauge fields become uncoupled; furthermore, in $L_{\rm dual}$ the gauge fields
appear only through the gauge field strengths $\Sigma$, with no explicit mention of the vector multiplet $V$.

The point now is to go to the limit $e' \rightarrow \infty$ on both theories, the original and the dual. For the original
$L$ it is well known that this limit takes us to the sigma-model with quotient target $\CC^n /\!/ T^{n-k} = X$ \cite{Hori-al}.
It is here that the Fano assumption on $X$ enters, as is also explained in \cite{Hori-al}. The resulting quotient model
is still gauged, with group $T^k$,  because we have kept the constant $e$ finite. On the dual side, on the other 
hand, the limit $e' \rightarrow \infty$ just imposes the constraints
\begin{equation}
Q^a_j\,  Y^j - t'^a \ = \ 0  \qquad {\rm for\ all}\quad a=1, \ldots , n-k.   \nonumber
\label{3.3}
\end{equation}
As in \cite{H-V}, these constraints are solved by 
\begin{equation}
Y^j \ = \ s^j \ +\ \sum^{k}_{b=1}  v^j_b \, \Theta^b    \ ,
\label{3.4}
\end{equation}
where the $\Theta^1 , \ldots , \Theta^k$ are new twisted chiral fields associated to new complex coordinates, and the
constants $s^1 , \ldots , s^n \in \CC$ are any particular solution of the algebraic equation $Q_j^a s^j = t'^a$. 
Finally, inserting (\ref{3.4}) into the limit of $L_{\rm dual}$ and denoting by $\langle \cdot , \cdot \rangle$ the 
canonical inner product on $\RR^k$, we get that as $e' \rightarrow \infty$ the lagrangian $L_{\rm dual}$ reduces to
\begin{align}
\hat{L}_{\rm dual} \ = &\ \int \dd^4 \theta\  \Big[ {\rm kinetic\ term}\ (\Theta^b , \bar{\Theta}^b)\  -\  \frac{1}{2 e^2}\:
\langle \bar{\Sigma},  \Sigma \rangle \Big]    \label{3.5}  \\
&+\, \frac{1}{2} \int \dd^2 \tilde{\theta}\ \bigg\{  \langle \Sigma , u_j  \rangle \Big(\langle v^j , \Theta \rangle  
+  s^j \Big)\; -\; \langle \Sigma  ,  t \rangle\;  +\;  \sum_{j=1}^{n} \exp{\big(- \langle v^j , \Theta \rangle - s^j\big)}  \ 
+\  {\rm c.c.} \bigg\} . \nonumber
\end{align}
This lagrangian is expected to be dual to the $e' \rightarrow \infty$ limit of $L$, i.e. dual to the lagrangian
of the gauged sigma-model with target $X$ and group $T^k$. The superpotential of this dual Landau-Ginzburg theory
is thus
\begin{equation}  
W \ = \ \langle \Sigma , u_j  \rangle \Big(\langle v^j , \Theta \rangle  +  s^j \Big) 
\ -\ \langle \Sigma  ,  t \rangle  \ +\  \sum_{j=1}^{n} \exp{\Big(- \langle v^j , \Theta \rangle - s^j\Big)} \ .
\label{3.6}
\end{equation}
We stress that these calculations are basically the same as in \cite{H-V}, the only difference being that here we have
kept the gauge terms in the quotient theory. In \cite{H-V} all the gauge terms were gotten rid of, because there one is only 
interested in the non-gauged sigma-model with target $X$.

\section{Observables of the dual topological theory} 

Our work up to this point has been to identify the Landau-Ginzburg theory dual to the original gauged sigma-model with
target $X$ and group $T^k$. We concluded that the dual theory has lagrangian (\ref{3.5}) and superpotential (\ref{3.6}).
We can now start to use this dual to extract properties of the gauged sigma-model.

The main object that concerns us in this note is the (small) quantum equivariant cohomology ring of the target $X$.
As is well known in the non-equivariant case, this ring coincides with, or can be defined as, the ring of local 
observables of the A-twisted topological sigma-model with target $X$. This topological theory is
obtained by twisting the original supersymmetric theory along its vectorial R-symmetry. Now the vectorial R-symmetry is 
also a symmetry of the dual model, the gauged Landau-Ginzburg model, and in particular one can also twist the latter
model to obtain a topological theory, which is then expected to be the dual of the gauged A-theory on $X$. The ring of
local observables of that topological Landau-Ginzburg model will then coincide with the ring of local observables of
the A-twisted gauged sigma-model, i.e. with the quantum equivariant cohomology ring of $X$. Thus, given this rationale,
we at present have the two following tasks: firstly to twist the dual theory (\ref{3.5}) along its vectorial R-symmetry, 
and, secondly, to compute the ring of observables of the resulting topological theory.

  The whole of this recipe is very familiar from the non-gauged case. In that case the final result is that the ring of 
local observables, or chiral ring, is a ring of polynomials quotiented by the ideal generated by the first partial 
derivatives of the superpotential $W$. Here in the gauged case, however, we must pay a little extra attention to the 
gauge fields. In fact, although the twisted chiral field strengths $\Sigma$ appear in the superpotential (\ref{3.6})
side by side with the twisted chiral matter fields $\Theta$, they are not exactly in the same footing as the latter. As we will see, at the end it turns out that the $\Sigma$'s do contribute to the ring of polynomials that corresponds to the observables but,
on the other hand, they do not contribute to the ideal by which one should mod-out, i.e. the partial derivatives of $W$
in the directions of the $\Sigma$'s do not appear in the result. The reason is that the observables determined by these partial derivatives are not trivial in the topological theory, and so cannot be moded-out. In contrast, the derivatives of $W$ with respect to the matter fields are still trivial (or $Q$-exact) observables, and hence as usual must be moded-out. This difference is explicitly revealed in the expressions for the action of the topological $Q$-operator obtained below in (\ref{4.2}).


$\ $

So we now want to determine the observables of the topological theory derived from the Landau-Ginzburg model
(\ref{3.5}). This model has the general form
\begin{equation}
\hat{L}\ = \ \int \dd^4 \theta\  \Big[ K (Y ,\bar{Y})\  -\  \frac{1}{2 e^2}\: \langle \bar{\Sigma},\Sigma \rangle \Big]\ 
+\ \frac{1}{2} \int \dd^2 \tilde{\theta}\ \Big\{ W( \Sigma , Y)\ +\ {\rm c.c.} \Big\} \ , 
\label{4.1}
\end{equation}
%
where the $\Sigma$'s and $Y$'s are the twisted chiral superfields of expressions (\ref{2.4}). As usual, after twisting, 
the space of fields of the topological theory will be acted by a natural fermionic operator, $Q$, and the observables of the 
theory will be the corresponding $Q$-cohomology classes. At the end of the section, after writing down the explicit form
of $Q$, we will conclude that the local observables of the topological theory are the holomorphic functions
$f (\sigma^b , y^j)$ modulo the ideal generated by the $Q$-exact functions $\partial_{y^j} W $. As mentioned before, the 
derivatives $\partial_{\sigma^b} W$ do not appear. This is, in fact, all that we will subsequently need, so the reader may 
wish at this point to just note the result and smoothly fly over to the next section. If not, then here is a brief 
justification. Starting with the lagrangian (\ref{4.1}) and superfields (\ref{2.4}), the first task is to twist the theory
along the vectorial R-symmetry. This symmetry is defined by
\begin{align}
Y\ &\longrightarrow \  Y (e^{-i\alpha} \theta^{\pm} , e^{i\alpha} \thetab^{\pm})  \nonumber \\
\Sigma \ &\longrightarrow \  \Sigma (e^{-i\alpha} \theta^{\pm} , e^{i\alpha} \thetab^{\pm})  \ ,\nonumber
\end{align} 
and in components reads
\begin{align}
\lamb_+ &\longrightarrow  e^{-i\alpha} \lamb_+   \qquad \qquad  \chib_+ \longrightarrow e^{-i\alpha} \chib_+    \nonumber \\
\lam_- &\longrightarrow  e^{i\alpha} \lam_-      \qquad \qquad  \chi_- \longrightarrow e^{i\alpha} \chi_-  \ , \nonumber 
\end{align}
with all the other fields remaining invariant. Applying the usual rules for twisting a theory along an R-symmetry 
\cite{Wit1, Hori-al},
we can reinterpret the various component fields as sections of different bundles and subsequently combine them into a new 
set of fields, the fields of the topological theory. These new fields will all be scalars or one-forms on the worldsheet
with values either on the Lie algebra or on the pull-back bundle $y^\ast TX$. The lagrangian of the topological theory 
is then obtained by writing down (\ref{4.1}) in components and substituting in the new set of fields. Now, because the fields
involved are many and we do not really need the explicit topological lagrangian for our purposes, we will skip writing it
down here. Instead we proceed directly to compute the form of the topological operator $Q$, which is the only
knowledge required to identify the ring of local observables. As usual, the action of $Q$ on the different fields can be 
read out of the $N=2$ supersymmetry transformations. These are written down for instance in \cite{H-V} for the gauge fields and in 
\cite{Hori-al} for the matter fields of the twisted chiral multiplet. 
Since the definition of the topological operator is $Q = Q_- + \bar{Q}_+$,
to obtain the explicit $Q$-action over the fields one only needs to put the fermionic parameters $\epsilon_+ = 
\bar{\epsilon}_- = 1$ and $\epsilon_- = \bar{\epsilon}_+ = 0$ in these transformations (see the conventions of \cite{H-V}). 
After integrating out the auxiliary fields $D^a$ and $G^j$ and rotating to euclidean worldsheet (because topological 
theories are euclidean), this yields the result:
\begin{align} 
Q\,& A^a_z \ =\ - i \lam_-^a   &   Q\,& \lam_-^a \ =\ -2\sqrt{2}\, \partial_z \sigma^a     \label{4.2}\\
Q\,& A^a_{\zb} \ =\  i \lamb_+^a   &   Q\,& \lamb_+^a \ =\  2\sqrt{2}\, \partial_{\zb} \sigma^a \nonumber  \\
Q\,& \sigma^a \ = \ 0  &  Q\,& (\lamb_-^a + \lam_+^a) \ = \ 0  \nonumber \\
Q\,& \sib^a \ = \ - i \sqrt{2} (\lamb_-^a + \lam_+^a)   &  Q\,&  (\lamb_-^a - \lam_+^a)  \ = \ -2\sqrt{2} i [ \ast F^a_A + 
 e^2  {\rm Re}\ ( \partial_{\sigma^{a}}W )]  \nonumber \\
Q\,& y^j \ = \ 0    &   Q\,& \chi_-^j \ =\ - 2 i \sqrt{2}\, \partial_z y^j  \nonumber  \\
Q\,& \bar{y}^j \ = \ \sqrt{2} (\chi^j_+ - \chib^j_- )    &   Q\,& \chib_+^j \ =\  2i \sqrt{2}\, \partial_{\zb} y^j \nonumber \\  
Q\,& (\chi^j_+ - \chib^j_- )\ =\ 0  &  Q\,& [2 \sqrt{2}\, h_{j\bar{l}}(\chi^l_+ + \chib^l_- )] \ =\ -\, \partial_{y^j} W \ .\nonumber 
\end{align}
Here $z$ is the complex coordinate on the euclidean worldsheet and $h_{j\bar{l}} = - \partial_j \partial_{\bar{l}}K$ 
is the hermitian metric on the geometry $T^k$-dual to $X$.
It is apparent from these expressions that the only natural $Q$-closed operators that may not be $Q$-exact are the 
holomorphic combinations of the fields $y$ and $\sigma$. Furthermore, one such holomorphic combination will be
$Q$-exact if and only if it has any of the derivatives $\partial_{y^j} W$ as a factor. Observe also that the partial
derivatives $\partial_{\sigma^b}W$ appear in (\ref{4.2}) only through their real part, which is not holomorphic and therefore 
not even $Q$-closed.


\section{The chiral ring of the gauged models}

Applying the results of the previous section to the lagrangian (\ref{3.5}), we see that the local observables of the dual 
topological theory are the holomorphic functions $f(\sigma^b , \theta^b)$ modulo the ideal of functions generated by the
partial derivatives $\partial_{\theta^b}W$. Now, in analogy to what is done in the non-gauged case, instead of considering 
all the holomorphic functions in the definition of chiral ring, we restrict ourselves to finite degree polynomials in the
variables $\sigma^b$ and $(x^b)^{\pm 1} := \exp{(\mp \theta^b)}$. This is related to the fact that in the definition of 
the equivariant de Rham complex we only consider finite degree forms and polynomials in the Lie algebra.
Then the chiral ring of the dual theory is
\begin{equation}
\CC [ \sigma^1, \ldots , \sigma^k , (x^1)^{\pm 1} , \ldots , (x^k)^{\pm 1} ] \ / \ D(W) \ ,
\label{5.1}
\end{equation}  
where $D(W)$ is the ideal generated by the derivatives
\begin{equation}
\partial_{\theta^b}W \ = \ -x^b\, \partial_{x^b} W \ = \ \sum_{j=1}^{n}\; u^b_j\: \Big[\langle \sigma , v^j \rangle \  -\ e^{-s^j} 
\prod_{c=1}^{k}  (x^c)^{v^j_c} \Big]\ . 
\label{5.2}
\end{equation}
This result can be cast in a different and perhaps simpler form. Consider the ring homomorphism
\begin{equation}
\Lambda :  \CC [\sigma^1 , \ldots , \sigma^k , w^1 , \ldots , w^n ]\ \longrightarrow \ \CC [\sigma^1 , \ldots , \sigma^k , 
(x^1)^{\pm 1} , \ldots , (x^{k})^{\pm 1} ]
\end{equation}
determined by
\begin{equation}
\sigma^b\  \longmapsto \ \sigma^c\, v^j_c \, u^b_j  \qquad {\rm and} \qquad w^j \ \longmapsto \ e^{-s^j}\: \prod_{c=1}^{k}\: (x^c)^{v^j_c} \ .
\end{equation}
It is not difficult to recognize that the matrix $B^b_c := u_j^b v_c^j$ is invertible over $\RR$. Basically this follows from
the fact that both $\{ Q^a , u^b  \}$ and $\{ Q^a , v_b \}$ are basis of $\RR^n$. It is then clear that, restricted to the 
subring of polynomials in the $\sigma$'s, the map $\Lambda$ is an isomorphism to its image, which is exactly the same subring.
Furthermore, because of the assumption that the vectors $v^1 , \ldots , v^n$ span the full $\mathbb{Z}^k$, it is also 
clear that the image by $\Lambda$ of the subring $\CC [ w^1 , \ldots , w^n ]$ is the full subring of Laurent polynomials 
$\CC [ (x^1)^{\pm 1} , \ldots , (x^{k})^{\pm 1} ]$. Hence, we conclude, $\Lambda$ is surjective. (Observe in passing that this 
was the first step that required the stated property of set of $v$'s, a property which comes from the regularity of the fan 
corresponding to $X$; in particular the derivation of (\ref{5.1}) did not require the regularity assumption on $X$.)

Now, by construction, the ideal $D(W)$ of (\ref{5.2}) is the image by $\Lambda$ of the ideal $\tilde{D}$ in the domain 
generated by the polynomials
\begin{equation}
\sigma^b - u_j^b \, w^j \qquad \qquad {\rm for} \qquad b = 1, \ldots , k \ .  \nonumber
\end{equation} 
Furthermore, just as in \cite{Bat}, it is apparent that the kernel of $\Lambda$ is the ideal $Q(X)$ generated by the binomials (\ref{1.2})
described in the introduction. Hence we finally conclude that, up to isomorphism of rings, 
\begin{align}
QH_T^\ast (X) \ &= \ \CC [\sigma^1 , \ldots , \sigma^k , w^1 , \ldots , w^n ]\ / \ (\tilde{D} + Q(X))  \label{5.4} \\
&=\ \CC [ w^1 , \ldots , w^n ]\ / \ Q(X) \ .  \nonumber
\end{align}

Having arrived at this result, a number of conclusions can be drawn in analogy with the non-equivariant case. Firstly observe that, through the definition (\ref{1.2}) of the ideal $Q(X)$, the chiral ring seems to depend on the complex parameter $t' = r' - i \theta'$ in $\CC^{n-k}$. It is easy to recognize, however, that this dependence is only apparent, and that by redefining the variables $w^j$ one can absorb any finite variation of $t'$. This means that the chiral ring of our gauged sigma-model does not depend on the value of $t'$, at least as long as the assumption that the target $X(r')$ is a smooth Fano manifold of dimension $k$ remains valid. This is analogous to the non-gauged case \cite{Bat, M-P}, and leads to the conclusion that toric targets that are isomorphic in codimension 1 have the same quantum equivariant ring. The usual example here is toric targets related by flop-type birational transformations, for in this case they can be realized as quotients $X(r')$ with the same charges $Q_j^a$ and parameters $r'$ belonging to adjacent cones in the parameter space ${\mathbb R}^{n-k}$.

Another conclusion follows from the result in \cite{Bat} that says that if $t'$ belongs to the K\"ahler cone of $X$ and $\lambda$ is a real scalar, then the formal limit $\lambda \rightarrow +\infty$ of the relations (\ref{1.2}) that define the ideal $Q_{\lambda t'}(X)$ is actually the set of relations that define the classical Stanley-Reisner ideal $J(X)$. Comparing $(\ref{1.1})$ and $(\ref{1.3})$ this shows that, as expected, the classical equivariant cohomology ring of $X$ can be obtained as a formal limit of the quantum equivariant cohomology ring. This formal limit, however, depends on the choice of the cone in $\CC^{n-k}$ to which $t'$ belongs. This corresponds to the fact that the classical equivariant ring of the quotient $X(r')$ also depends on the cone, in contrast with the quantum equivariant ring.

We thus conclude that much of the story of mirror symmetry and quantum cohomology of toric manifolds extends to the equivariant case.  

$\ $

\noindent {\bf Example: projective spaces.} 
The compact toric manifolds $X = \CC {\mathbb P}^k$ correspond to the assignments $n=k+1$ and $Q= (1, \ldots , 1) 
\in {\mathbb Z}^{k+1}$ in the general picture above. The result (\ref{5.4}) then states that the quantum equivariant
cohomology ring of these manifolds is the polynomial ring
\begin{equation*}
Q H_{T^k}^\ast (\CC {\mathbb P}^k) \ = \ \CC [w^1 ,  \ldots , w^{k+1}] /  (w^1 \cdots w^{k+1} - e^{-t'}) \ .
\end{equation*}
The parameter space is here one-dimensional, and in the limit $t' \rightarrow +\infty$ this ring formally reduces to the classical $T^k$-equivariant cohomology
of $\CC {\mathbb P}^k$, which is
\begin{equation*}
 H_{T^k}^\ast (\CC {\mathbb P}^k) \ = \ \CC [w^1 ,  \ldots , w^{k+1}] /  (w^1 \cdots w^{k+1}) \ . 
\end{equation*}
To put this in context, recall that the usual (non-equivariant) cohomology rings of $\CC {\mathbb P}^k$ are
\begin{equation*}
Q H^\ast (\CC {\mathbb P}^k) \ = \ \CC [w] /  (w^{k+1} - e^{-t'}) \qquad \ {\rm and} \qquad \ 
H^\ast (\CC {\mathbb P}^k) \ = \ \CC [w] /  (w^{k+1}) \ . \
\end{equation*}

\vskip 45pt
\noindent
{\bf Acknowledgements.}
I would like to thank Marcos Mari\~no and Jan de Boer for helpful conversations and the referee for suggesting several improvements.  I am partially supported by the 
Netherlands Organisation for Scientific Research (NWO), Veni grant 639.031.616.

\end{document}